\documentclass[aps,prl,reprint,showpacs,superscriptaddress]{revtex4-1}
\usepackage{graphicx}% Include figure files
\usepackage{dcolumn}% Align table columns on decimal point
\usepackage{bm}% bold math
\usepackage{color}

\begin{document}
%\preprint{}

%%  For including Chinese characters, try one of the following:
%\begin{CJK*}{GB}{gbsn}
%%\begin{CJK*}{GB}{}
%% To insert Chinese characters, use any of the Chinese character soft keyboards, such as ѹ

\title{Suppression of transverse ablative Rayleigh-Taylor-like instability 
       in the hole-boring radiation pressure acceleration 
       by using elliptically polarized laser pulses}
% Please use the appropriate macro for each type of information
%\thanks{}
\author{D. Wu}
\affiliation{Key Laboratory of HEDP of the Ministry of Education, CAPT, and State Key Laboratory of Nuclear Physics and Technology, Peking University, Beijing, 100871, China.}
\author{C. Y. Zheng}
\email{zheng\_chunyang@iapcm.ac.cn}
\affiliation{Key Laboratory of HEDP of the Ministry of Education, CAPT, and State Key Laboratory of Nuclear Physics and Technology, Peking University, Beijing, 100871, China.}
\affiliation{Institute of Applied Physics and Computational Mathematics, Beijing, 100088, China.}
\author{B. Qiao}
\affiliation{Key Laboratory of HEDP of the Ministry of Education, CAPT, and State Key Laboratory of Nuclear Physics and Technology, Peking University, Beijing, 100871, China.}
\author{C. T. Zhou}
\affiliation{Key Laboratory of HEDP of the Ministry of Education, CAPT, and State Key Laboratory of Nuclear Physics and Technology, Peking University, Beijing, 100871, China.}
\affiliation{Institute of Applied Physics and Computational Mathematics, Beijing, 100088, China.}
\author{X. Q. Yan}
\affiliation{Key Laboratory of HEDP of the Ministry of Education, CAPT, and State Key Laboratory of Nuclear Physics and Technology, Peking University, Beijing, 100871, China.}
\author{M. Y. Yu}
\affiliation{Institute of Fusion Theory and Simulation, Zhejiang University, Hangzhou, 310027, China.}
\author{X. T. He}
\email{xthe@iapcm.ac.cn}
\affiliation{Key Laboratory of HEDP of the Ministry of Education, CAPT, and State Key Laboratory of Nuclear Physics and Technology, Peking University, Beijing, 100871, China.}
\affiliation{Institute of Applied Physics and Computational Mathematics, Beijing, 100088, China.}
\date{\today}
\begin{abstract}
It is shown that the transverse Rayleigh-Taylor-like (RT) instability in the hole boring radiation pressure acceleration
can be suppressed by using elliptically polarized (EP) laser. 
A moderate $\bm{J}\times\bm{B}$ heating of the EP laser will thermalize the local electrons, which leads to the transverse diffusion of ions, suppressing the short wavelength perturbations of RT instability.  
A proper condition of polarization ratio is obtained analytically for the given laser intensity and plasma density. The idea is confirmed by two dimensional Particle-in-Cell simulations, showing that the ion beam driven by the EP laser is more concentrated and intense compared with that of the circularly polarized laser.  
\end{abstract}
\pacs{52.38.Kd, 41.75.Jv, 52.35.Mw, 52.59.-f}
\maketitle
%\end{CJK*}

%%Also, the American journals encourage Asian authors to include their original names. So I have also included the Chinese names. If you cannot compile this latex file, you should try to see if the other  begin\{CJK*\}\{GB\}\{gbsn\} line works.}

\textit{Introduction.}{---}Recently, ion acceleration from the interaction of ultra-intense laser pulse with plasmas has attracted wide attention because of its broad applications, including producing high energy density matter, ion-fast ignition in laser fusion, tumor therapy and 
radio-graphing \cite{PhysRevLett.106.145002, PhysRevLett.86.436, PhysRevLett.89.175003, PhysRevLett.88.215006}. Almost all of these applications call for a high quality ion beam with large particle number, sharp energy spread and low divergence angle. Radiation pressure acceleration (RPA) is a potential scheme for generating high quality ion beams, including (light sail) LS-RPA for thin foil \cite{Rev.Mod.Phys.85.751, PhysRevLett.103.024801, PhysRevLett.102.145002, PhysRevLett.105.155002, PhysRevLett.105.065002, PhysRevLett.105.065002,
PhysRevLett.100.135003, PhysRevLett.103.135001, PhyPla.16.044501, PhysRevLett.108.225002}
and (hole boring) HB-RPA for thick target \cite{PhyPla.18.056701, PhysRevLett.102.025002, PhyPla.16.083130, PlaPhyConFus.51.024004, PlaPhyConFus.51.095006, PhyPla.14.073101, PhyPla.18.073101, PhyPla.16.033102, PhysRevLett.106.014801}. In particular, the HB-RPA owns the intrinsic property for large particle number acceleration \cite{PhyPla.18.053108}. In HB-RPA, the ponderomotive force drives the local electrons inward, resulting in a shock-like double layer (DL) region with large electrostatic charge separation field. The latter could reflect the ions initially located ahead of the DL, accelerating them like a piston. For an efficient long laser pulse and thick target,
the reflection will repeat and generate a large number of particles. For a usual circularly polarized (CP) laser driven HB-RPA, the DL oscillations would broad the energy spread of the accelerated ion beams \cite{PlaPhyConFus.51.024004}. The present authors \cite{PhyPla.20.023012} proposed to use elliptically polarized (EP) laser to suppress the DL oscillations, generating high quality mono-energetic ion beams compared with that of CP laser. However the crucial concern of the RPA acceleration stability is the transverse Rayleigh-Taylor-like (RT) instability \cite{PhysRevLett.105.065002,PhyPla.18.073106,PhysRevLett.108.225002}. The classical RT instability can occur when a light fluid pushes or accelerates a heavy fluid, and this situation is very similar to the RPA case, where the photons act as light fluid and plasmas as heavy fluid \cite{PhysRevLett.108.225002}. 
The RT instability will break the target surface and terminate the acceleration process. 

\begin{figure}
\includegraphics[width=6.00cm]{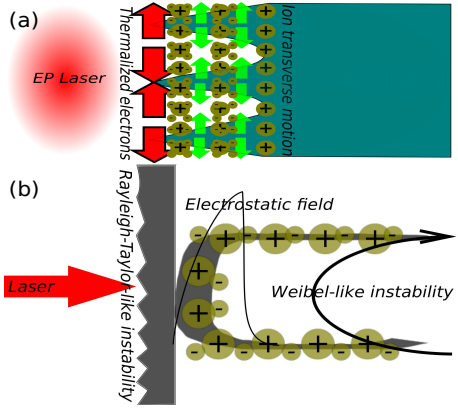}
\caption{\label{f1} (color online) 
(a) The suppression mechanism of the EP laser driven ablative RT instability. The EP lasers moderately heat electrons, providing the transverse diffusion of the ions.
(b) The schematic demonstration of different instabilities in the HB-RPA process in the piston-rest-frame. RT instability takes place on the laser plasma interface, and Weibel like or filamentation instability and two-stream are driven by the incident and reflected particle beams.}
\end{figure}

In this Letter, we propose to use EP laser for stabilization of the transverse RT instability in the HB-RPA. The idea is similar to the stabilization mechanism for ablative RT instability in the inertial confinement fusion (ICF) research. Unlike the classical RT instability, the short wavelength ablative RT instability in ICF research can be suppressed due to thermal smoothing of the perturbations \cite{Phys.Fluids.28.3676,PhyPla.3.1402,PhyPla.113.690,PhyPla.3.2122,PhysRevLett.98.245001}.  
As is shown in Fig.\ \ref{f1} (a), because of the moderate $\bm{J}\times\bm{B}$ effect, EP laser will thermalize those electrons located within the DL region to high temperature. The high electron temperature provides a fast transverse diffusion of ions with velocity equal to sound speed $({T_eZ/m_i})^{1/2}$,  
where $T_e$ is the electron temperature, $Z$ is the ion charge number and $m_i$ is the ion mass. The transverse diffusion of the ions could suppress the transverse ablative RT instability efficiently. During the time of the RT instability growth, the diffusion range of the ions could overshoot the instability wavelength if a fast diffusion velocity is generated. To have faster ion diffusion velocity, a higher electron temperature or smaller polarization ratio $\alpha=a_z/a_y$ of EP laser is needed. On the other hand, as expected, if $\alpha$ is too small ($\alpha=0$ for linearly polarized laser), the laser piston structure is totally destroyed. Thus, there should be a lower limit for the polarization ratio $\alpha$ to sustain the HB-RPA process. Based on these ideas, a proper condition of polarization ratio is obtained analytically for the given laser intensity and plasma density. The theory is confirmed by two dimensional (2D) Particle-in-Cell (PIC) simulations. 

\begin{figure}
\includegraphics[width=8.5cm]{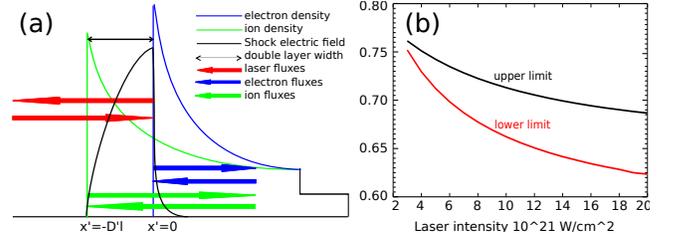}
\caption{\label{f2} (color online) (a) Schematic structure of the hole-boring radiation acceleration process in the piston-rest-frame.
(b) The proper range of polarization ratio $\alpha$ for the given hydrogen plasmas of density $n=20$ vs. laser intensity. }
\end{figure}

\textit{Theoretical model.}{---}In the HB-RPA regime as shown in Fig.\ \ref{f2} (a) \cite{PhysRevLett.102.025002, PhyPla.16.083130}, the ponderomotive force drives the local electrons inward, resulting in a shock like DL region with large electrostatic charge separation field. The latter traps and reflects the ions.
We consider in the piston-rest-frame. 
For an EP laser pulse of amplitude $a_y(=eE_{y}/m_e\omega_{0}c)$ and $a_z(=eE_{z}/m_e\omega_{0}c)$, the propagating velocity of the piston $c\beta_f$ can be obtained from the momentum balance equation, $2I_{0}(1-\beta_f)/(1+\beta_f)=2n_i(m_i+Zm_e)c^3\gamma_f^2\beta_f^2$, where $I_0=(a_y^2+a_z^2)n_cm_ec^3/2$ is the EP laser intensity, $n_i$ is the ion density, $Z$ is the ion charge number, $m_i$ and $m_e$ are ion and electron mass.
In the piston-rest-frame with the propagating velocity $\beta_f=[{I_{0}/n_i(m_i+Zm_e)c^{3}}]^{1/2}/$ $[1+[{I_{0}/n_i(m_i+Zm_e)c^{3}}]^{1/2}]$, 
the radiation pressure of the EP laser reads $p_{rad}=2I_0(1-\beta_f)/(1+\beta_f)/c$,
where the term $(1-\beta_f)/(1+\beta_f)$ is the modification of the laser frequency due to the Doppler effect, 
and full laser reflection is assumed.

Because of the smaller mass of electrons, the accelerated ion beam is accompanied with an electron beam, keeping almost quasi-neutral. If we think of the motion of ion and electron as a whole,
it is reasonable to neglect the electron inertia, and assume that the ponderomotive force acts on the ions directly.
As the ion acceleration is limited within the DL region, which is about tenth of laser wavelength \cite{PhysRevLett.29.1429,PhysRevLett.99.065002,PhyPla.19.043104}, the mass density of the thin layer is 
$\sigma_m =\int_{-D^{\prime}_l}^{0}m_i n^{\prime}_idx^{\prime} \approx m_i n_i c/\omega_{pi}$,
where $D^{\prime}_l$ is the DL width, $n_i^{\prime}$ is the ion density distribution within the DL region and $\omega_{pi}=({4\pi Ze^{2}n_e/m_i})^{1/2}$ is the characteristic frequency of ions.
{As the ions keep on entering into and being reflected from the DL, the density and laser intensity distribution therein do not change with time, we can treat this DL as a static layer. 
The acceleration $g$ of the laser radiation pressure acted on this DL can be expressed as 
$g=p_{rad}/\sigma_m$. Similar to the RT instability in the LS-RPA regime \cite{PhysRevLett.99.065002}, where a thin target is driven by high intensity laser, RT instability would also dominate the transverse behavior of this DL.}

We consider two points $(x_0,y_0)$ and $(x_0,y_0+\delta y_0)$ on
this DL layer. These two points will evolve at some time to the points $(x,y)$ 
and $(x+\delta y_0\partial{x}/\partial{y_0},y+\delta y_0\partial{y}/\partial{y_0})$. The $x$ and $y$ components of the force equation of the element on this DL in the accelerating frame can be written as
\begin{eqnarray}
\label{Eq:equation-of-motion}
&&\partial{p_x}/\partial{t}=-g\sigma_m d y_0+p_{rad}dy_0\partial{y}/\partial{y_0}, \\
&&\partial{p_y}/\partial{t}=-p_{rad}dy_0\partial{x}/\partial{y_0},
\end{eqnarray}
where $p_x=\gamma_f \sigma_m d y_0 dx/dt$, $p_y=\gamma_f \sigma_m d y_0 dy/dt$ and $-g$ is the inertial acceleration. After some arrangement, 
Eqs.\ (\ref{Eq:equation-of-motion}) and (2) can be rewritten as
\begin{eqnarray}
\label{Eq:equation-of-motion-simplified}
&&\partial^{2}{x}/\partial{t}^{2}=-g/\gamma_f+(\partial{y}/\partial{y_0})g/\gamma_f, \\
&&\partial^{2}{y}/\partial{t}^{2}=-(\partial{x}/\partial{y_0})g/\gamma_f.
\end{eqnarray}
The solution of Eqs.\ (\ref{Eq:equation-of-motion-simplified}) and (4) turns out to be \cite{PhysRevLett.29.1429}
\begin{eqnarray}
\label{Eq:equation-of-motion-solution}
&&x=\delta_0 \exp[t(kg/\gamma_f)^{1/2}]\cos(ky_0), \\
&&y=y_0-\delta_0 \exp[t(kg/\gamma_f)^{1/2}]\sin(ky_0),
\end{eqnarray}
where $\delta_0$ is the initial disturbed amplitude, $k=2\pi/\lambda_{rt}$, and $\lambda_{rt}$ is the instability wavelength.
If the disturbed amplitude in the $x$ direction is over $1/k$ at sometime $\tau^{'}$, it means that the adjacent sections of this DL begin to collide with each other \cite{PhysRevLett.29.1429}. Here the transverse RT instability has already developed, and we define $\tau^{'}$ as the characteristic time of the linear RT instability. From Eq.\ (\ref{Eq:equation-of-motion-solution}), we have $\tau^{'}=\log(1/\delta_0k)(\gamma_f/kg)^{1/2}$.
After transforming to the laboratory-frame, $\tau=\gamma_f\tau^{'}$, the characteristic time of the linear RT instability reads,
\begin{eqnarray}
\label{Eq:equation-of-motion-solution}
\tau=\log(1/\delta_0k)\frac{\lambda^{1/2}\gamma^{3/2}_f (m/Z)^{3/4}n^{1/4}(1+\beta_f)^{1/2}}{2\pi(a_y^2+a_z^2)^{1/2}(1-\beta_f)^{1/2}},
\end{eqnarray}
where $m=m_i/m_e$, $n=n_e/n_c$, $n_c$ is the critical density, $\lambda=\lambda_{rt}/\lambda_0$ is the instability wavelength normalized to laser wavelength, and $\tau$ is normalized to laser period $T_0$.

It should be emphasized the characteristic time $\tau$ is not sensitive to $\delta_0$ and $k$, due to the slow varying of $\log$ function. 
The factor $\log(1/\delta_0k)$, to some degree, is constant, which can be determined from the PIC simulations.
Once the value of $\log(1/\delta_0 k)$ is chosen, it will keep constant for all cases of different laser intensity and plasma density for our analysis.  
According to 2D PIC simulations, the constant factor is defined as $\log(1/\delta_0k)=13.8$. This equation demonstrates that the shorter wavelength perturbations grow more faster compared with longer wavelength perturbations, and that stronger laser intensity and lower plasma density correspond to faster instability growth rate.
This instability growth rate is qualitatively in agreement with that obtained by F. Pegoraro $\textit{et al.}$ considering the LS-RPA cases \cite{PhysRevLett.99.065002}. 
 
When an EP laser irradiates the target, its moderate oscillating ponderomotive force will heat the local electrons. Considering the electron gamma factor under the EP laser, 
$\gamma_e=[{1+[a_y\cos(\omega_0t)]^2+[a_z\sin(\omega_0t)]^2}]^{1/2}$,
and assuming $1+a_y^2+a_z^2>>a_y^2-a_z^2$, the gamma factor can be transfered to, using Taylor expanding formula,
$\gamma_e=\gamma_{0}+\gamma_{2}\cos(2\omega_0t)$,
where $\gamma_{0}=({1+a_y^2/2+a_z^2/2})^{1/2}$ is the constant gamma term and $\gamma_{2}=(a_y^2-a_z^2)/4\gamma_{0}$ is the oscillating gamma term.
The thermalization of the local electrons is dominated by the time oscillating term, in which the temperature of the local electrons can be approximated to $m_ec^2(\gamma_{2}-1)$. Thus, the transverse diffusion velocity of ions, i.e. the sound speed, is $v_d=[{(\gamma_{2}-1)Z/m}]^{1/2}$, where $v_d$ is normalized to the light speed $c$.

As the schematic Fig.\ \ref{f1} (a) shows, in order to suppress the transverse ablative RT instability, during the characteristic time $\tau$, the diffusion range of the ions should overshoot the perturbation wavelength \cite{PhyPla.18.073106}. To smooth the short wavelength perturbations with wavelength  $\lambda\sim1$, we should make sure that the condition $\tau v_d > 1$ is satisfied. 
After some arrangement (setting $a_y=a_0$ and $a_z=\alpha a_0$), this condition equally reads
\begin{eqnarray}\label{14}
\label{Eq:upper-limit}
&& a_0^2(1-\alpha^2)/4[{1+a_0^2/2(1+\alpha^2)}]^{1/2}-1 \nonumber \\ 
&& > \eta a_0^2(1+\alpha^2)(1-\beta_f)/(m/Z)^{1/2}\gamma_f^3n^{1/2}(1+\beta_f),
\end{eqnarray}
where $\eta=[2\pi/\log{(1/\delta_0k)}]^{2}=0.21$.
For the given laser intensity $I_0=a_0^2(1+\alpha^2)n_cm_ec^3/2$ and plasma condition $m$, $Z$ and $n$, coupled with Eq.\ (\ref{Eq:upper-limit}) we can solve the polarization ratio $\alpha$ out. 
The obtained polarization ratio is the upper limit to ensure that the electron temperature is high enough to thermally smooth RT instability.

There should be a lower limit of polarization ratio to keep the HB-RPA process. For an usual laser piston structure shown in Fig.\ \ref{f2} (a), the $\bm{J}\times\bm{B}$ effect of EP laser will drag the local electrons forward to vacuum.
The lower limit of polarization ratio is derived based on the assumption that these forward-going electrons are stopped within the DL region to keep the acceleration structure intact. Balancing the (forward) kinetic and the electrostatic and ponderomotive potential energies of the electrons, we can obtain the lower limit of polarization ratio $\alpha$ \cite{PhyPla.20.023012}, 
\begin{eqnarray}
\label{Eq:lower-limit}
&& m(\gamma_f-1)/Z+[{1+a_0^4(1-\alpha^2)^2/16}]^{1/2}-1 \nonumber \\ 
&& < m^{1/2}a^2_0(1+\alpha^2)\beta_f(1-\beta_f)/3 Z^{1/2}(1+\beta_f)\gamma_f n.
\end{eqnarray}
Similar to Eq.\ (\ref{Eq:lower-limit}), for the given laser intensity and plasma condition, $\alpha$ can also be solved out. The corresponding polarization ratio is the lower limit to ensure the HB-RPA acceleration process.  
Inequalities (\ref{Eq:upper-limit}) and (\ref{Eq:lower-limit}) determine two polarization ratio conditions. 
For hydrogen plasma with density $n_e=20n_c$, 
a range of polarization ratio vs. laser intensity is shown in Fig.\ \ref{f2} (b).

\begin{figure}
\includegraphics[width=8.50cm]{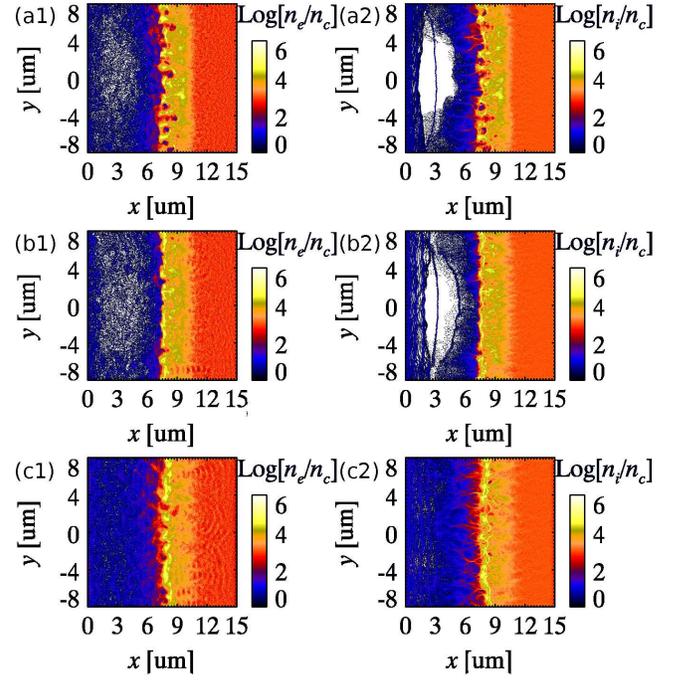}
\caption{\label{f3} (color online) The density distribution of electrons (column $1$) and protons (column $2$) at $t=40T_0$.
(a), (b) and (c) correspond to CP laser, EP laser of $\alpha=0.7$ and EP laser of $\alpha=0.3$. Here the initial hydrogen plasma density is of $n_e=20n_c$, and laser intensity is of $6.85\times10^{21}$ W$/$cm${^2}$ with wavelength $1.0 \mu$m.}
\end{figure}

\textit{Numerical results.}{---}2D PIC (\texttt{KLAP} code \cite{PhysRevLett.107.265002}) simulations are carried out to confirm this theory. The size of the simulation box is $L_y\times L_x=16\lambda_0(y)\times15\lambda_0(x)$ with $\lambda_0$ representing the laser wavelength. The simulation box is divided into uniform grid of $1600(y)\times1500(x)$. The laser pulse enters into the simulation box from the left boundary. 
The bulk target consists of two species: electrons and protons, which are initially located in the region $-8.0\lambda_0<y<8.0\lambda_0$ and $2.0\lambda_0<x<15.0\lambda_0$ with density $n_e=20n_c$, where $n_c=\omega_0^2e^2m_e/4\pi=1.1\times10^{21}$ /cm$^3$ is the critical density for $1.0$ $\mu$m laser pulse. We use $100$ particles per cell to run the simulations. The initial plasma temperature is set to be $1.0$ keV. The normalized amplitude of CP laser is $a_y=50.00$ and $a_z=50.00$, corresponding to the laser intensity $6.85\times10^{21}$ W$/$cm${^2}$. 
The laser temporal profile is of Gaussian with the half-width-half-maximum pulse duration $35T_0$. 
In contrast,  EP laser pulses with the same space and temporal profile are run. Following Fig.\ \ref{f2} (b), for laser intensity of $6.85\times10^{21}$ W$/$cm${^2}$, polarization ratio of $\alpha=0.7$ ($\alpha=0.3$) is chosen, which is within (out of) the proper range.

\begin{figure}
\includegraphics[width=8.5cm]{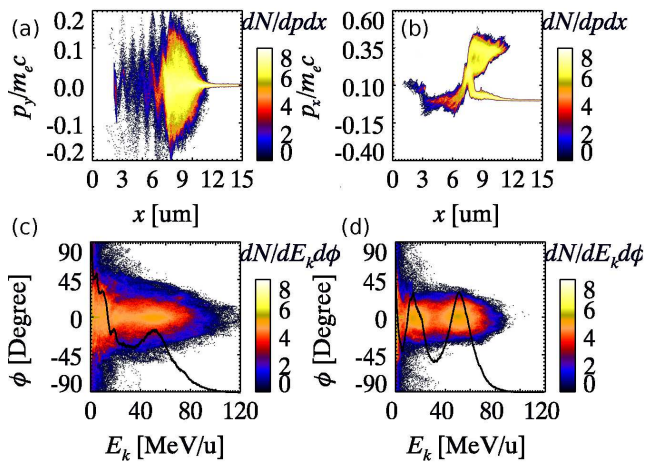}
\caption{\label{f4} (color online) 
(a) The $(x,p_y)$ phase diagram of protons driven by EP laser ($\alpha=0.7$) at $t=40T_0$.
(b) The $(x,p_x)$ phase diagram of protons driven by EP laser ($\alpha=0.7$) at $t=40T_0$.
(c) and (d) The angular distributions and energy spectrum (over the contour plot) of the proton beams driven by CP and EP ($\alpha=0.7$) lasers at $t=40T_0$, respectively.} 
\end{figure} 

The results are shown in Fig.\ \ref{f3}. When driven by CP laser, the ponderomotive force has no second-harmonic oscillating component, and there is no suppression effect at all. As shown in Fig.\ \ref{f3} (a) the transverse RT instability has developed at $t=40T_0$, with the perturbation wavelength about $1.0\lambda_0$. The characteristic time of RT instability from PIC simulation is consistent with prediction of theoretical model, which is $\tau\sim30T_0$ obtained from Eq.\ (\ref{Eq:equation-of-motion-solution}). 
The instability pattern is similar to that reported by F. Pegoraro $\textit{et al.}$ \cite{PhysRevLett.99.065002}. These short wavelength perturbations can severely disturb the hole boring process.  
In contrast, when driven by EP laser ($\alpha=0.7$), due to the $\bm{J}\times\bm{B}$ effect, the EP laser will thermalize those electrons located within the DL region. The effective electron heating provides a fast transverse diffusion of the ions with the sound speed $({T_eZ/m_i})^{1/2}$.
Following our model, the temperature of the heated electrons can be as high as $3$ MeV, which means that the diffusion velocity can be as fast as $v_d=0.05$. The diffusion velocity from theoretical model is also in agreement with that obtained from PIC simulation as shown in Fig.\ \ref{f4} (a). It is this transverse diffusion of the ions that suppresses the transverse ablative RT instability, which is clearly shown in Fig.\ \ref{f3} (a) and (b). 
During the characteristic time of the RT instability, the diffusion range of the ions can be as far as $v_d \tau\sim1.5\lambda_0$, well overshooting the instability wavelength $1.0\lambda_0$.  
If the polarization ratio is out of the proper range, such as $\alpha=0.3$ in Fig.\ \ref{f3} (c), although the RT instability is also thermally suppressed, the over heating of the EP laser would destroy the HB-RPA process, and large amount of electrons on the piston surface are dragged out into the backward vacuum.  
{From Fig.\ \ref{f4} (b), we can see the ions are reflected smoothly by the shock electric field.} 
The energy spectrum is shown in Fig.\ \ref{f4} (c) and (d), where both total particle number and energy spread of the proton beam driven by the EP laser ($\alpha=0.7$) are much better than that of the CP laser, because of the suppression of RT instability.

Figure\ \ref{f4} (c) and (d) also shows the angular distribution of the accelerated proton beams driven by CP and EP lasers respectively. Compared with (c) and (d), we see that the transverse diffusion of protons driven by EP laser does not have a obvious effect on the beam angular divergence. For EP laser, the transverse diffusion velocity of protons is about $v_d=0.05$, while the longitudinal velocity is double the piston forward velocity, which can be as high as $v_l=0.35$. 
The divergence angle can be approximated to be $\tan^{-1}(v_d/v_l)\sim8.0$ degree. It is reasonable to say that the small divergence angle can still be held under the mechanism we proposed. 
It is demonstrated in Fig.\ \ref{f4} (d) that the divergence angles of proton beams driven by EP laser can be maintained within $10.0$ degree, which is also consistent with the predictions of theory.

\textit{Discussions.}{---}On the instabilities of HB-RPA, except for the RT instability, the filamentation and two-stream instabilities might also plan important roles during the particle propagation process \cite{PhyPla.18.073101,PhyPla.19.103105}. We have found that these different types of instabilities dominate in different regions of the target. As Fig.\ \ref{f1} (b) shows, for RT instability, it is driven on the laser plasma interface, while filamentation and two-stream instabilities are induced in the inner part of the target during the propagation process of the accelerated particles. The RT instability can severely disturb the laser plasma interface and strongly affect the quality of the accelerated ions. When RT instability builds up at a high level, the laser plasma interface is totally destroyed and the HB-RPA is also terminated. {Compared with filamentation and two-stream instabilities, RT instability is the key to stabilize and elongate the HB-RPA process. The $\bm{J}\times\bm{B}$ heating of the EP laser could enhance the background temperature of the electrons, and affect the Weibel like instabilities, however our previous theoretical studies show that a high temperature background can reduce the Weibel like instabilities to some degree \cite{PhyPla.20.032113}. 
Compared with ideal CP laser, EP laser is easier to realize in real experiment.}
For a through understanding of the effect of EP laser on RT, filamentation and two-stream instabilities, it is beyond the contents of this work, and shall be studied in the following works.

{In our simulation, the plasma density is chosen to be $20n_c$ only to confirm our theory. However, unlike LS-RPA, the underlying physics (the velocity of the piston and the thermal smoothing mechanism) only dependents on normalized laser amplitude $a=eE/m_e\omega_0c$ for HB-RPA. By using long laser wavelength, such as $10$ $\mu$m, the laser intensity and plasma density are correspondingly reduced by two orders of magnitude, that is to say the laser intensity is now $6.85\times10^{19}$ W$/$cm${^2}$ and the plasma density is $2.2\times10^{20}$ /cm$^3$. Our theory can be verified in experiment by CO$_2$ laser irradiating gas target.} 

\textit{Conclusions.}{---}In summary, we propose to use EP laser to suppress the transverse RT instability in the 
hole boring radiation pressure acceleration. The moderate $\bm{J}\times\bm{B}$ effect of the EP laser will thermalize the local electrons, inducing the transverse diffusion of ions and resulting in the stabilization of the short wavelength
perturbations of RT instability. 
The proper condition of polarization ratio is obtained analytically for the given laser intensity and plasma density. The theory is confirmed by 2D PIC simulations. The obtained ion beam driven by EP laser is more concentrated and intense compared with that driven by CP laser. The beam divergence is not severely affected by the transverse smoothing mechanism. Relatively smaller divergence angle within $10.0$ degree can be maintained in the proposed scheme. 
\begin{acknowledgments}
This work was supported by the National Natural Science Foundation of China (Grant Nos. 11075025), 
and National Basic Research Program of China (grant No. 2013CB834100).
\end{acknowledgments}

%\bibliography{reference}

\begin{thebibliography}{99}
\bibitem{PhysRevLett.106.145002} C. Wang, X.-T. He, and P. Zhang, Phys. Rev. Lett. 106, 145002 (2011).

\bibitem{PhysRevLett.86.436} M. Roth, T. E. Cowan, M. H. Key, S. P. Hatchett, C. Brown, W. Fountain, J. Johnson, D. M.
Pennington, R. A. Snavely, S. C. Wilks, K. Yasuike, H. Ruhl, F. Pegoraro, S. V. Bulanov,
E. M. Campbell, M. D. Perry, and H. Powell, Phys. Rev. Lett. 86, 436 (2001).

\bibitem{PhysRevLett.89.175003} T. Z. Esirkepov, S. V. Bulanov, K. Nishihara, T. Tajima, F. Pegoraro, V. S. Khoroshkov,
K. Mima, H. Daido, Y. Kato, Y. Kitagawa, K. Nagai, and S. Sakabe, Phys. Rev. Lett. 89, 175003 (2002).

\bibitem{PhysRevLett.88.215006} A. J. Mackinnon, Y. Sentoku, P. K. Patel, D. W. Price, S. Hatchett, M. H. Key, C. Andersen,
R. Snavely, and R. R. Freeman, Phys. Rev. Lett. 88, 215006 (2002).

\bibitem{Rev.Mod.Phys.85.751} Andrea Macchi, Marco Borghesi and Matteo Passoni, Rev. Mod. Phys. 85, 751 (2013).

\bibitem{PhysRevLett.103.024801} M. Chen, A. Pukhov, T. P. Yu, and Z. M. Sheng, Phys. Rev. Lett. 103, 024801 (2009).

\bibitem{PhysRevLett.102.145002} B. Qiao, M. Zepf, M. Borghesi, and M. Geissler, Phys. Rev. Lett. 102, 145002 (2009).

\bibitem{PhysRevLett.105.155002} B. Qiao, M. Zepf, M. Borghesi, B. Dromey, M. Geissler, A. Karmakar, and P. Gibbon, 
Phys. Rev. Lett. 105, 155002 (2010).

\bibitem{PhysRevLett.105.065002} T.-P. Yu, A. Pukhov, G. Shvets, and M. Chen, 
Phys. Rev. Lett. 105, 065002 (2010).

\bibitem{PhysRevLett.100.135003} X. Q. Yan, C. Lin, Z. M. Sheng, Z. Y. Guo, B. C. Liu, Y. R. Lu, J. X. Fang, and J. E. Chen,
Phys. Rev. Lett. 100, 135003 (2008).

\bibitem{PhysRevLett.103.135001} X. Q. Yan, H. C. Wu, Z. M. Sheng, J. E. Chen, and J. Meyer-ter Vehn, Phys. Rev. Lett. 103,
135001 (2009).

\bibitem{PhyPla.16.044501} X. Q. Yan, M. Chen, Z. M. Sheng, and J. E. Chen, 
Phys. Plasmas 16, 044501 (2009).

\bibitem{PhysRevLett.108.225002} C. A. J. Palmer, J. Schreiber, S. R. Nagel, N. P. Dover, C. Bellei, F. N. Beg, S. Bott, 
R. J. Clarke, A. E. Dangor, S. M. Hassan, P. Hilz, D. Jung, S. Kneip, S. P. D. Mangles, K. L. Lancaster, 
A. Rehman, A. P. L. Robinson, C. Spindloe, J. Szerypo, M. Tatarakis, M. Yeung, M. Zepf, and Z. Najmudin1, 
Phys. Rev. Lett. 108, 225002 (2012).

\bibitem{PhyPla.18.056701} A. P. L. Robinson, Phys. Plasmas 18, 056701 (2011).

\bibitem{PhysRevLett.102.025002} N. Naumova, T. Schlegel, V. T. Tikhonchuk, C. Labaune, I. V. Sokolov, and G. Mourou,
Phys. Rev. Lett. 102, 025002 (2009).

\bibitem{PhyPla.16.083130} T. Schlegel, N. Naumova, V. T. Tikhonchuk, C. Labaune, S. I. V., and G. Mourou, 
Phys. Plasmas 16, 083103 (2009).

\bibitem{PlaPhyConFus.51.024004} A. P. L. Robinson, P. Gibbon, M. Zepf, S. Kar, R. G. Evans, and C. Bellei, 
Plasma Phys. and Controlled Fusion 51, 024004 (2009).

\bibitem{PlaPhyConFus.51.095006} A. P. L. Robinson, D.-H. Kwon, and K. Lancaster, 
Plasma Phys. and Controlled Fusion 51, 095006 (2009).

\bibitem{PhyPla.14.073101} X. M. Zhang, B. F. Shen, X. M. Li, Z. Y. Jin, and F. C. Wang, Phys. Plasmas 14, 073101 (2007).

\bibitem{PhyPla.18.073101} Xiaomei Zhang, Baifei Shen, Liangliang Ji, Wenpeng Wang, Jiancai Xu, Yahong Yu, and Xiaofeng Wang,
Phys. Plasmas 18, 073101 (2011).

\bibitem{PhyPla.16.033102} Xiaomei Zhang, Baifei Shen, Zhangying Jin, Fengchao Wang, and Liangliang Ji,
Phys. Plasmas 16, 033102 (2009).

\bibitem{PhysRevLett.106.014801} Charlotte A. J. Palmer, N. P. Dover, I. Pogorelsky, M. Babzien, G. I. Dudnikova, M. Ispiriyan, M. N. Polyanskiy, J. Schreiber, P. Shkolnikov, V. Yakimenko, and Z. Najmudin,
Phys. Rev. Lett. 106, 014801 (2011).

\bibitem{PhyPla.18.053108} J. Badziak, G. Mishra, N. K. Gupta, and A. R. Holkundkar,
Phys. Plasmas 18, 053108 (2011).

\bibitem{PhyPla.20.023012} Dong Wu, C. Y. Zheng, C. T. Zhou, X. Q. Yan, M. Y. Yu, and X. T. He,
Phys. Plasmas 20, 023012 (2013).

\bibitem{PhyPla.18.073106} Min Chen, Naveen Kumar, Alexander Pukhov, and Tong-Pu Yu, Phys. Plasmas 18, 073106 (2011).

\bibitem{Phys.Fluids.28.3676} H. Takabe, K. Mima, L. Montierth, and R. L. Morse, Phys. Fluids 28, 3676 (1985).

\bibitem{PhyPla.3.1402} V. N. Goncharov, R. Betti, R. L. McCrory, P. Sorotokin, and C. P. Verdon,
Phys. Plasmas 3, 1402 (1996).
 
\bibitem{PhyPla.113.690}P. Clavin and L. Masse, Phys. Plasmas 11, 690 (2004).

\bibitem{PhyPla.3.2122} R. Betti, V. N. Goncharov, R. L. McCrory, P. Sorotokin, and C. P. Verdon, Phys. Plasmas 3, 2122 (1996).

\bibitem{PhysRevLett.98.245001} R. Masse, Phys. Rev. Lett. 98, 245001 (2007).

\bibitem{PhysRevLett.29.1429} Edward Ott, Phys. Rev. Lett. 29, 1429 (1972).

\bibitem{PhysRevLett.99.065002} F. Pegoraro and S. V. Bulanov, Phys. Rev. Lett. 99, 065002 (2007).

\bibitem{PhyPla.19.043104} Ujjwal Sinha, Phys. Plasmas 19, 043104 (2012).

\bibitem{PlaPhyConFus.51.024014} V K Tripathi , C S Liu , X Shao , B Eliasson and R Z Sagdeev,
Plasma Phys. and Controlled Fusion 51, 024014 (2009).

\bibitem{PhysRevLett.107.265002} H. Y. Wang, C. Lin, Z. M. Sheng, B. Liu, S. Zhao, Z. Y. Guo, 
Y. R. Lu, X. T. He, J. E. Chen and X. Q. Yan, 
Phys. Rev. Lett. 107, 265002 (2011).

\bibitem{Nat.Phys.8.95} Dan Haberberger, Sergei Tochitsky, Frederico Fiuza, Chao Gong, Ricardo A. Fonseca,
Luis O. Silva, Warren B. Mori and Chan Joshi, Nat. Phys. 8, 95, (2011). 

\bibitem{PhyPla.19.103105} S. V. Bulanov, T. Zh. Esirkepov, M. Kando, F. Pegoraro, S. Bulanov, C. G. R. Geddes, 
C. B. Schroeder, E. Esarey, and W. P. Leemans, Phys. Plasmas 19, 103105 (2012). 

\bibitem{PhyPla.20.032113} Qing Jia, Hong-bo Cai, Wei-wu Wang, Shao-ping Zhu, Z. M. Sheng, and X. T. He, Phys. Plasmas 20, 032113 (2013). 
\end{thebibliography}
{}

\end{document}